\documentclass[prb,preprint,superscriptaddress,onecolumn,aps]{revtex4-2}
\usepackage{amsmath}  % needed for \tfrac, \bmatrix, etc.
\usepackage{amsfonts} % needed for bold Greek, Frater, and blackboard bold
\usepackage{graphicx} % needed for figures
\usepackage{url}
\usepackage{natbib}
\begin{document}
\title{Inexpensive Benchtop Soft Photolithography Technique for Microfluidics and Other Applications}
\author{Habilou Ouro-Koura}
\affiliation{Department of Mechanical Engineering, Rensselaer Polytechnic Institute, Troy, NY 12180 USA}
\affiliation{University of Maryland Eastern Shore, 1, Backbone Road, Princess Anne, MD 21853 USA}
\author{Ayobami Ogunmolasuyi}
\affiliation{Department of Mechanical Engineering, Dartmouth College, Troy, NY 12180 USA}
\affiliation{University of Maryland Eastern Shore, 1, Backbone Road, Princess Anne, MD 21853 USA}
\author{Othman Suleiman}
%\author{XX XX}
\author{Kausik S Das}
\email{kdas@umes.edu} % optional
\affiliation{University of Maryland Eastern Shore, 1, Backbone Road, Princess Anne, MD 21853 USA}
\affiliation{Kavli Institute of Theoretical Physics, Santa Barbara, CA, 93106 - 4030 USA}
%Rensselaer Polytechnic Institute, Troy, NY 12180 USA}

%\email{hourokoura@umes.edu}

%\date{\today}

\begin{abstract}
\noindent  Microfluidics have emerged as a vital tool in various scientific and engineering applications, ranging from biomedical diagnostics to chemical analysis. However, conventional fabrication methods often necessitate costly equipment and limiting accessibility, thereby hindering widespread adoption. In response, this note presents a cost-effective approach for fabricating microfluidic channels using photolithographic methods, which circumvents the need for expensive instruments.

\noindent The procedure commences with the preparation of a suitable substrate, followed by the application of a photoresist layer. A photomask, containing the desired microfluidic pattern, is then used to define the layout on the photoresist-coated substrate. Upon exposure to ultraviolet (UV) light, the photoresist undergoes a chemical transformation, allowing for selective removal of either exposed or unexposed regions. Consequently, the patterned microfluidic channels are unveiled on the substrate.

\noindent The versatility of this photolithographic method allows for the fabrication of intricate and customizable microchannels with relative ease and precision. Further processing steps, such as etching or surface treatment, optimize the substrate's properties for specific applications. This low cost approach not only empowers researchers and engineers to develop microfluidic devices at a fraction of the cost but also fosters broader access to photolithography related research and development.

\noindent By eliminating the financial barrier associated with expensive instruments, this cost-effective fabrication technique holds significant potential to revolutionize access to photolithography based research even in undergraduate laboratories. The enhanced accessibility opens doors to a myriad of applications, ranging from advanced diagnostics and drug delivery systems to portable point-of-care devices. Overall, this paper showcases a promising avenue for expanding the reach of microfluidics, fostering innovation, and driving impactful advancements in diverse scientific disciplines.
\end{abstract}

\maketitle
\section{Introduction}
Microfluidics involves the precise manipulation of small volumes of fluids within channels ranging from tens to hundreds of micrometers in dimensions, allowing for the study of novel phenomena occurring at the microscale to nanoscale. This field revolves around controlling and guiding fluids within channels with micrometer-scale or smaller dimensions, where capillary action and diffusion and boundary conditions play dominant roles in transport mechanisms.

Microfluidics has found widespread applications in various scientific and engineering domains, including low-cost nanoparticle synthesis\cite{chen2022intelligent}, DNA nanostar condensation\cite{conrad2022emulsion}, cell biology research\cite{paguirigan2008microfluidics}, drug delivery\cite{riahi2015microfluidics}, oil recovery\cite{fani2022application, foroughi2012immiscible} fuel cells\cite{kjeang2009microfluidic, das2020microbial}, drops and bubble dynamics\cite{kenning2006confined, das2010dynamics} controlled ion transport\cite{liu2020ion, nazari2020surface} and lab-on-a-chip investigations\cite{stone2004engineering}. One of its major advantages lies in its ability to enable controlled and precise analysis while using minimal volumes of samples, chemicals, and reagents, thus reducing overall costs and promoting sustainability. Additionally, its compact size facilitates parallel processing, enabling multiple operations to be executed simultaneously, thereby shortening experimental timelines.

Our laboratory has harnessed the concept of benchtop soft photolithography to design micro-channels in polydimethylsiloxane (PDMS) microfluidic structures, allowing for micro-mixing experiments within straight channels. The work\cite{barnes2021plasma} comprises experimental procedures to develop and test these micro-channels, perform fluid experiments using the lab-made microfluidic setups, and subsequently validate the results through numerical simulations of fluid flow within the channels.

This experimental methods paper provides a detailed account of how to fabricate PDMS microchannels using soft lithography techniques, highlighting the significance of microfluidics in advancing research and applications in diverse fields. By harnessing these cost-effective methods, our lab aims to contribute to the continued growth and impact of microfluidics technology in various scientific and engineering endeavors.

\section{Equipment and supplies}
\begin{figure}[ht!]
\centering
\includegraphics[angle=0,width=7.0in]{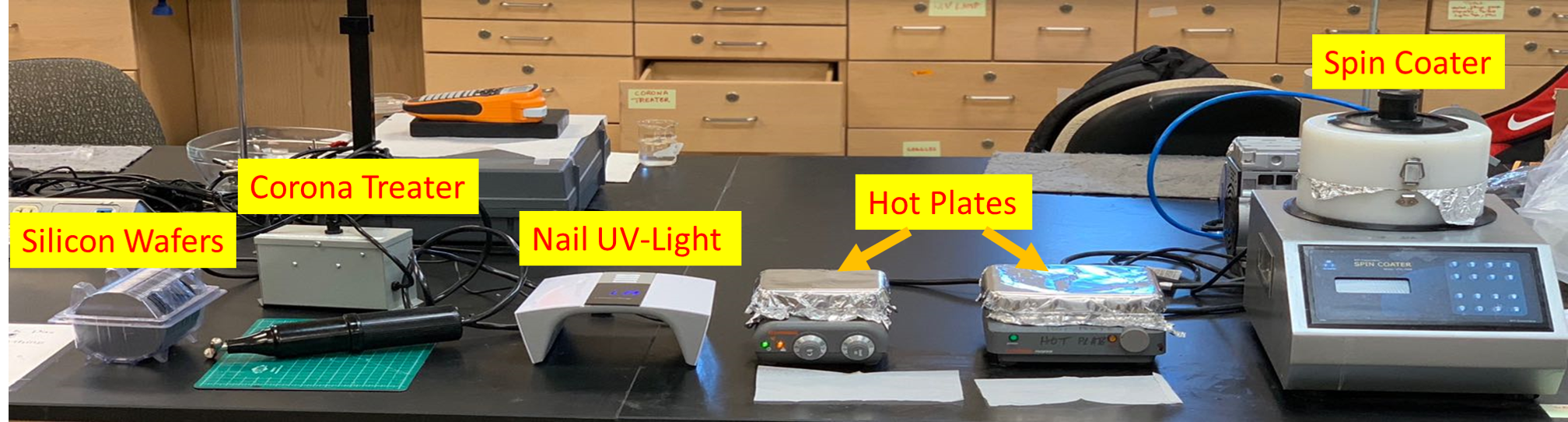}
\caption{Materials and equipment needed for benchtop photolithography.}
\label{materials}
\end{figure}
\begin{itemize}
  \item Silicon Wafer. University Wafers.
  \item Spin Coater. MTI Corporation.
  \item Hot Plate. Fischer Scientific.
  \item Corona Treater. Electro-Technic Products, BD-20AC Laboratory Corona Treater.
  \item Low cost photomask printed on transparency, CAD Art Services Inc. www.outputcity.com
  \item UV lamp. Piano Series 007, UV LED Nail Lamp (Adjustable 48W/ 60W)
  \item Incubator. Quincy Lab, Inc. , Model 10-140 Incubator, 140 Series
  \item Microfluidic pump. New Era Pump Systems, Inc. Model No. NE-300
  \item Biopsy Puncher. Electron Microscopy Science Rapid-Core 0.75,
  \item L-shaped metal connector, TE Needle 20 GA, Bent 90 Degree, Amazon
  \item Tubing. 1.6mm OD, 0.8mm ID.
\end{itemize}

\section{Chemicals Used}
\begin{itemize}
  \item MicroChem SU-8 2050 photoresist
  \item MicroChem SU-8 Developer
  \item Acetone, Carolina Biological
  \item Isopropyl Alcohol, Carolina Biological
  \item Deionized Water.
\end{itemize}
\section{Pre-processing and setup}
Estimated time for completion: 2 hours (Note: Use the polished face of the wafer for the entire process). \newline
Steps: \newline
 1) Activate two hot plates, setting one at 65°C and the other at 95°C.\newline
 2) Prepare a plastic disposable glass containing 10mL of Micro-Chem SU-8 developer. \newline
 3) Thoroughly clean all tabletop apparatus used with Isopropyl Alcohol (IPA).\newline
 4) Take precaution to prevent accumulation of dust and contamination on the wafer during the entire process.\newline

\section{Wafer Preparation Procedure}
\begin{figure}[ht!]
\centering
\includegraphics[angle=0,width=3.2in]{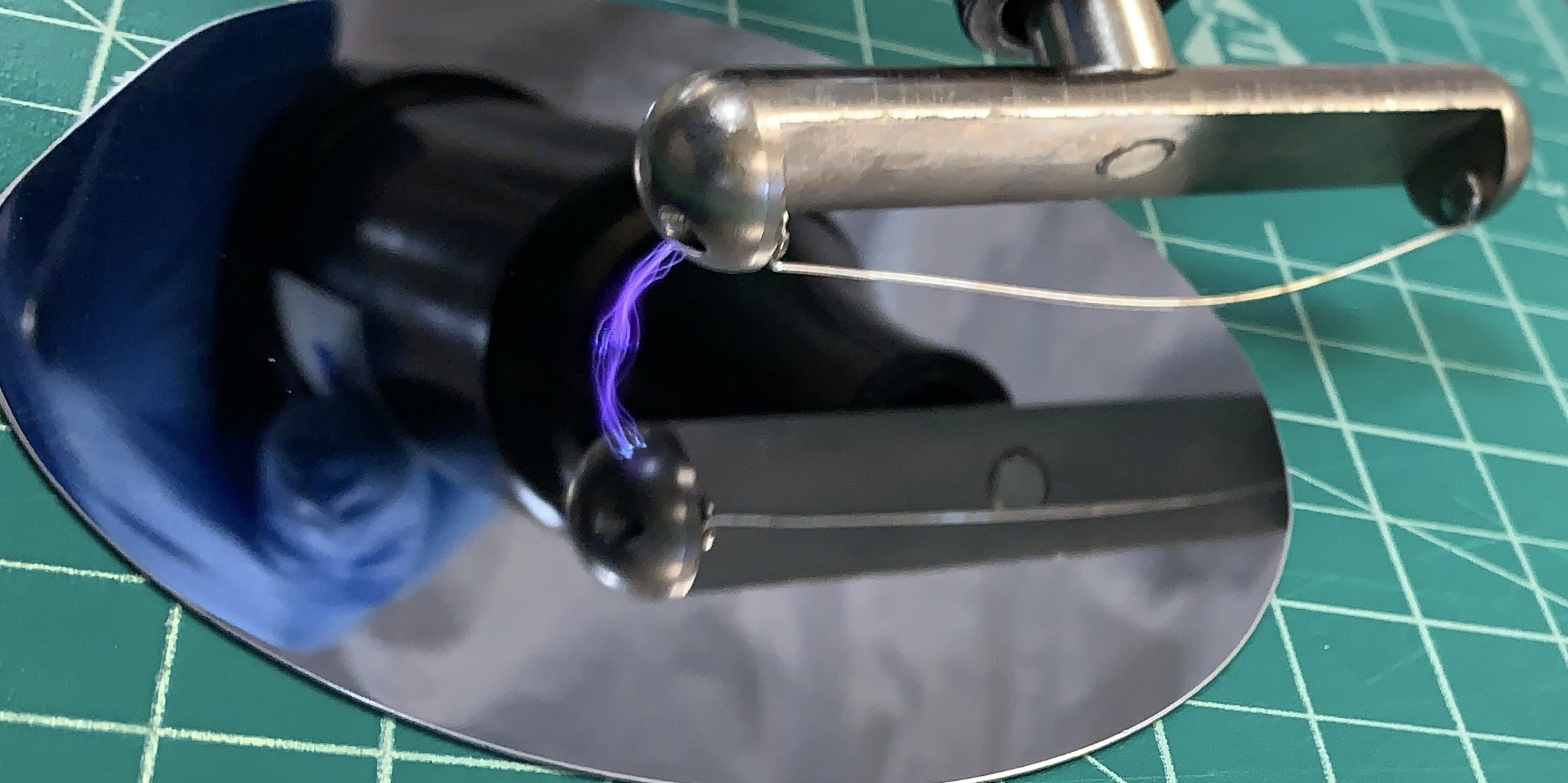}
\caption{Silicon wafer treatment with corona discharge.}
\label{corona1}
\end{figure}

In order to prepare the Silicon Wafer for further processing, follow the following steps:

\begin{enumerate}
  \item \textbf{Cleaning the Wafer:} Start by cleaning the Silicon Wafer with Acetone followed by Isopropyl Alcohol (IPA). Gently wipe the surface of the wafer with a lint-free cloth (such as eye glass cleaning cloth) or swab soaked in Acetone, removing any residues or contaminants. Then repeat the same procedure with Isopropyl Alcohol to ensure a thorough cleaning.

  \item \textbf{Rinsing with Deionized (DI) Water:} Once the cleaning is completed, rinse the entire wafer properly with Deionized (DI) water. This step is crucial to remove any remaining traces of solvents and contaminants.

  \item \textbf{Drying the Wafer:} To dry the wafer, use compressed air with low air speed to avoid any mechanical stress or damage to the wafer. Ensure that the wafer surface is completely dry before proceeding to the next step.

  \item \textbf{Baking the Wafer:} Place the wafer on a hot plate and set the temperature to 95$^\circ$C. Let the wafer bake for approximately 10 minutes. This step is essential to evaporate any remaining water on the surface of the wafer, ensuring optimal conditions for subsequent processes.

  \item \textbf{Cooling Down:} After the baking process, carefully remove the wafer from the hot plate and allow it to cool down to room temperature. This cooling period usually takes around 5 minutes, and it ensures the wafer reaches a stable temperature for subsequent treatments.

  \item \textbf{Treating with Corona Treater/Plasma:} Utilize a handheld corona treater to treat the entire surface of the wafer for 20 seconds. The corona treater generates a controlled discharge, enhancing surface wettability and promoting adhesion for subsequent coatings. It also helps breaking down any remaining organic matters on the surface of the wafer. Alternatively, if you have a plasma cleaner in your lab, plasma treat the wafer for a few seconds.

  \item \textbf{Applying Photoresist (PR):} Immediately after treatment with the corona treater/plasma cleaner, spin coat the photoresist (PR) on the wafer. Use a suitable spin coater to ensure an even and uniform distribution of the photoresist across the wafer surface.

\end{enumerate}

Once all these steps are completed, the silicon wafer is now ready for the desired microfabrication or processing steps.

%\section{Spin coating of the negative photoresist}
% In our case SU-8 2050 photoresist was used for the fabrication of the microchannels. At first load the wafer on the vacuum chuck of the spin coater. Align the wafer properly so that the center of the wafer rests on the middle of the chuck. Then, pour 2-4mL of the PR, depending on the wafer size (usually bigger diameter wafer requires more PR to coat its surface) on the middle of the wafer. While pouring the PR avoid trapping air bubbles by pouring it as a continuous stream. Use the following coating recipe to coat the wafer with the PR.
%
%Step 1: Speed (500 rpm); Acceleration/Deceleration: 300rpm/s; Duration:10 sec, followed by
%
%Step 2: Speed: 2000 rpm; Acceleration/Deceleration: 500rpm/s; Duration: 30sec.

\section{Spin Coating of the Negative Photoresist}

For the fabrication of microchannels, we have used the SU-8 2050 negative photoresist. The spin coating process is a crucial step in preparing the wafer for subsequent photolithographic processes. Follow these instructions to perform the spin coating:

\begin{enumerate}
  \item \textbf{Wafer Loading:} Place the Silicon Wafer on the vacuum chuck of the spin coater. Ensure proper alignment, positioning the center of the wafer at the middle of the chuck.

  \item \textbf{Pouring the Photoresist:} Pour 2-4mL of the negative photoresist (PR) onto the middle of the wafer. The amount of PR required depends on the wafer size; larger diameter wafers may need more PR to coat their surfaces effectively. Take care to avoid trapping air bubbles during this step, as they can affect the coating quality.

  \item \textbf{Coating Recipe:} Use the following coating recipe to coat the wafer with the photoresist:

    \textbf{Step 1:}
    \begin{itemize}
      \item Speed: 500 rpm
      \item Acceleration/Deceleration: 300 rpm/s
      \item Duration: 10 seconds
    \end{itemize}

    \textbf{Step 2:}
    \begin{itemize}
      \item Speed: 2000 rpm
      \item Acceleration/Deceleration: 500 rpm/s
      \item Duration: 30 seconds
    \end{itemize}

\end{enumerate}

The spin coating process ensures a uniform and controlled distribution of the negative photoresist over the wafer surface. Once the spin coating is complete, the wafer is ready for the photolithography step.

\section{Soft Pre-exposure-baking}

The soft pre-exposure-baking process is designed to gradually heat up the photoresist (PR). Rapid heating can lead to stresses and cracks in the PR film. Follow these steps to perform the soft pre-exposure-baking:

\begin{enumerate}
  \item \textbf{Wafer Removal:} Carefully remove the wafer from the spin coater after the spin coating process.

  \item \textbf{First Hot Plate:} Place the wafer on the first hot plate set at 65$^\circ$C. Allow the wafer to undergo pre-exposure baking at this temperature for 10 minutes. This gentle heating process helps to enhance the adhesion and stability of the PR film.

  \item \textbf{Second Hot Plate:} Transfer the wafer to the second hot plate set at 95$^\circ$C. Continue the pre-exposure baking at this temperature for 30 minutes. This extended baking step further improves the PR's mechanical properties, reducing the likelihood of deformation during subsequent processing.

  \item \textbf{Avoid Direct Light Exposure:} After the pre-exposure baking is complete, ensure that the coated wafer is not exposed to bright direct light. Direct light exposure can induce photochemical reactions in the PR, leading to undesired changes in the photoresist properties.

  \item \textbf{Cooling Down:} Once the baking steps are finished, allow the wafer to cool down to room temperature for approximately 5 minutes. The cooling period ensures that the wafer reaches a stable temperature, making it ready for the photolithography process.

\end{enumerate}

The soft pre-exposure-baking process optimizes the PR film's quality and stability, ensuring better performance during subsequent photolithography steps.

\section{Designing and Printing of Inexpensive Photomasks for Microfluidic Channels}

The fabrication of microfluidic channels using photolithography requires the design and printing of cost-effective photomasks. Photomasks act as templates defining the desired patterns and structures on the photoresist-coated wafer. Follow these steps to create affordable photomasks:

\begin{enumerate}
  \item \textbf{Designing the Photomask:} Utilize computer-aided design (CAD) software or specialized photomask design software to create the layout of the microfluidic channels. Ensure precision and accuracy in defining the desired channel dimensions and geometries.

  \item \textbf{Choose Substrate:} To keep the cost low, select an economical substrate material for the photomask. Common options include glass plates, plastic transparencies, or even overhead projector sheets.

  \item \textbf{Photomask Printing:} Use a high-resolution printer with a transparency printing option to print the photomask pattern onto the selected substrate. Adjust the printer settings to achieve the desired resolution and transparency density suitable for photolithography. Alternatively, there are several vendors, such as CAD Art Services Inc. etc., who provide high resolution photomask printing service on transparency films or chrome.

  \item \textbf{Verify Accuracy:} After printing, verify the accuracy of the photomask pattern using a microscope or any imaging tool to ensure that the channels' design is correctly replicated.

  \item \textbf{Cleanliness:} Ensure the photomask's surface is free from any dust or debris that might affect the photolithography process. Handle the photomask with care to avoid scratches or damage.

  \item \textbf{Mask Alignment:} During the photolithography process, accurately align the photomask with the photoresist-coated wafer to ensure proper channel formation.

  \item \textbf{Reuse or Dispose:} Depending on the material and cleanliness, inexpensive photomasks can be reused for multiple runs of photolithography or disposed of after a single use.

\end{enumerate}

By following these steps mentioned avove and leveraging affordable materials and printing methods, you can create cost-effective photomasks that facilitate the fabrication of microfluidic channels, making this advanced technology more accessible to researchers in schools, colleges, primarily undergrad institutions and industries alike.

\begin{figure}[ht!]
\centering
\includegraphics[angle=0,width=5.0in]{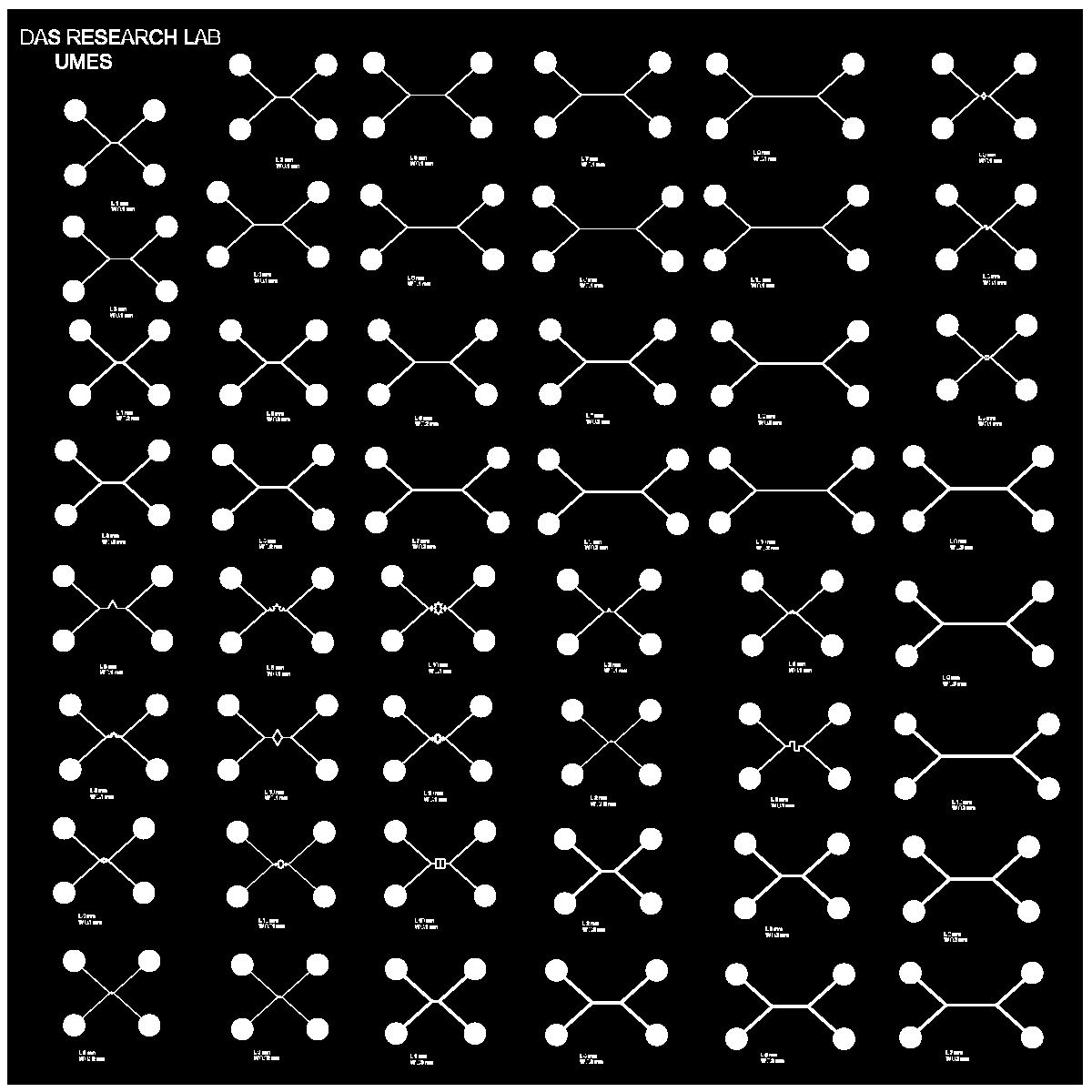}
\caption{Undergrad student designed photomask for microchannels of different length, width and shapes printed on a transparency film.}
\label{Photomask}
\end{figure}

\section{Exposure to UV light}
\begin{figure}[ht!]
\centering
\includegraphics[angle=0,width=3.2in]{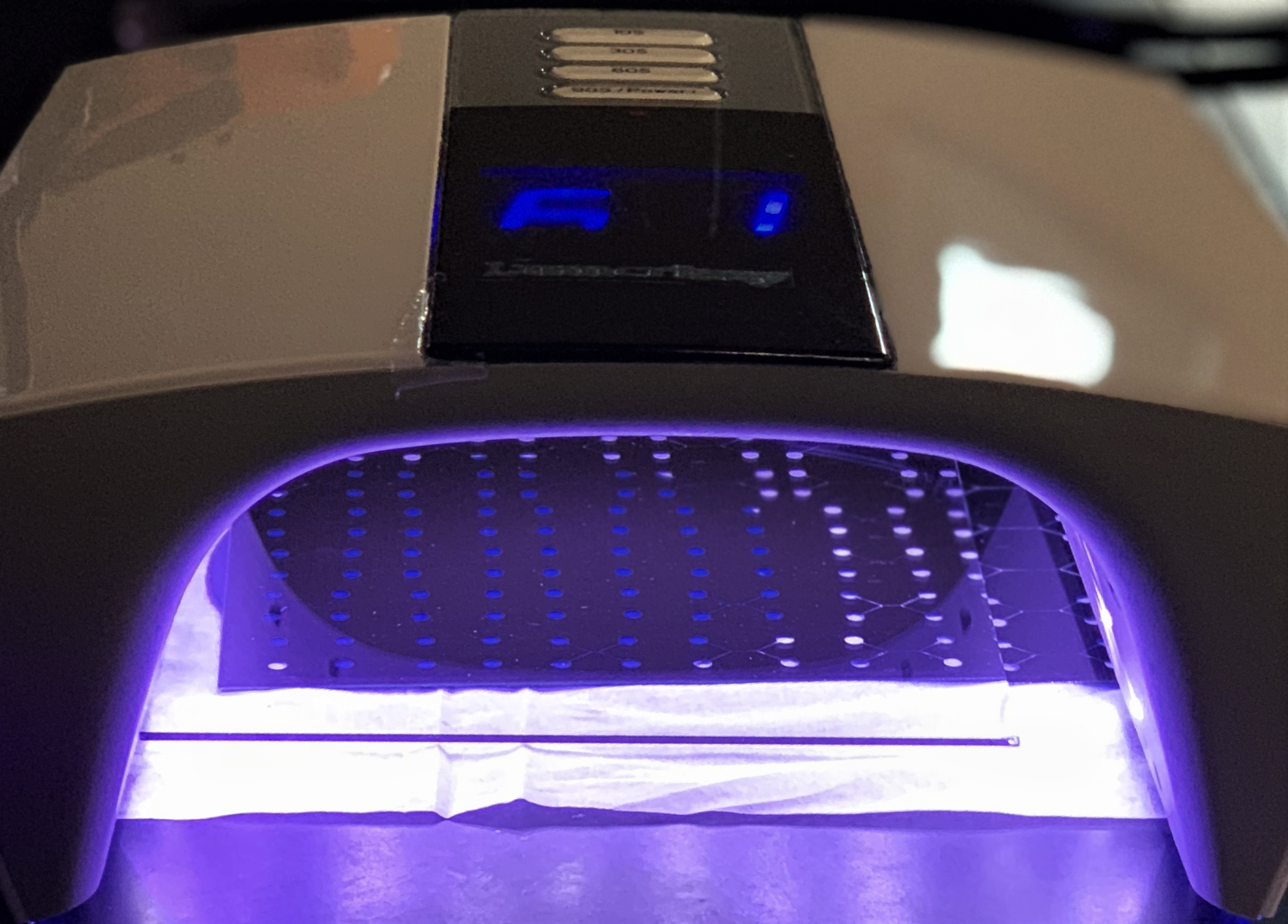}
\caption{UV treatment of the coated wafer using a nail curer.}
\label{UV}
\end{figure}

Before turning on the UV light nail curer (365 nm wavelength), position the wafer underneath it. Ensure precise alignment of the photomask (PM), which contains the design of the microchannels, on the wafer. To enhance contact between the photomask and the Photoresist-coated wafer, place a wide flat glass plate on top of the photomask.

Activate the UV light at maximum power and expose the wafer-photomask combination to the UV light for a duration of 5 minutes. After exposure, allow the setup to stand for an additional 5 minutes.

With caution, remove the glass slide and the Photomask from the wafer, completing the photolithography step.

\section{Development}
To begin the development process, place the wafer in the MicroChem SU-8 developer solution kept in a tray. Gently swirl the developer solution for 5-10 minutes or until the unexposed photoresist dissolves completely in the developer. As soon as the microchannels become visible, promptly remove the wafer from the developer using a tweezer and rinse it with isopropyl alcohol (IPA). Note that if there are any white traces on the wafer, it indicates that the photoresist was not entirely developed. In such cases, re-immerse the wafer into the developer for 30 seconds to 1 minute, and repeat the process until no white traces remain during rinsing with IPA. Once the development process is complete, rinse the microchannels with deionized water and dry them slowly with compressed air. Finally, bake the wafer on a 65$^\circ$C hot plate for 5 minutes. The microchannels (molds) are now ready for further processing.

\section{PDMS Mold for Microfluidic Chip}
For creating microchannels, we will be using SYLGARD 184 elastomer. Here are the steps:

\begin{enumerate}
  \item Open the SYLGARD 184 kit, which contains two plastic bottles. The larger bottle contains the PDMS base, while the smaller bottle contains the PDMS curing agent.
  \item Place a disposable plastic container on a scale.
  \item Carefully pour the PDMS base into the container.
  \item Next, add the PDMS curing agent into the container, maintaining a weighing ratio of 1 to 10. For instance, if the PDMS base is 10g, you will need 1g of the PDMS curing agent.
  \item Thoroughly mix the PDMS mixture in the container to ensure a homogenous blend. Take your time, don't rush.
  \item Now, place the silicon wafer with the microchannel molds in a plastic container. We have used plastic petri dish whose diameter is slightly more than the wafer diameter. 
  \item Carefully pour the PDMS mixture over the silicon wafer, ensuring that there is enough PDMS to cover the wafer, leaving at least 0.5 cm of PDMS covering its surface.

\end{enumerate}

The PDMS mold with microchannels is now prepared, ready for the subsequent steps in the microfluidic chip fabrication process.

%
%\section{Degassing of PDMS (getting rid of the air bubbles)}
%It is important to get rid of the air bubbles in the PDMS mixture.
%There are two methods to get rid of the air bubbles:
%- Degassing using the vacuum pump for 2 hours or more if needed. (This is the quickest
%way to take out the air bubbles)
%- Leaving the PDMS mix uncovered for 24 hours at room temperature.
%For this process, the second method is used.
%Leave the container containing the silicon wafer and PDMS in the lab without covering and at
%room temperature for 24 hours.
%After the 24 hours, there will be no more air bubbles in the PDMS mix. The PDMS will be still soft
%and sticky to touch.
%Place the container in the incubator at 40$^\circ$C for 1 hour or more until the PDMS is no more sticky to touch.

\section{Degassing of PDMS (Getting Rid of Air Bubbles)}
It is crucial to eliminate air bubbles from the poured PDMS mixture to ensure optimal results in the fabrication process. There are two methods for degassing the PDMS:

\begin{itemize}
  \item \textbf{Method 1: Vacuum Pump Degassing (Quick Method):} Use a vacuum pump to degas the PDMS mixture for 2 hours or more, if necessary. This method efficiently removes air bubbles and is the quickest way to achieve bubble-free PDMS.

  \item \textbf{Method 2: Uncovered Degassing (24 Hours Method):} For this process, the second method is used. Leave the container containing the silicon wafer and PDMS in the lab without covering it and at room temperature for 24 hours.

\end{itemize}

After the 24-hour period, the PDMS mixture should be free of air bubbles. However, it will still be soft and sticky to the touch. To further prepare the PDMS for use:

\begin{enumerate}
  \item Place the container with the silicon wafer and PDMS in the incubator at 40$^\circ$C for 1 hour or more until the PDMS is no longer sticky to the touch.

\end{enumerate}

Following these steps ensures that the PDMS is degassed and ready for use in creating the microfluidic chip.

\section{Cutting the PDMS Microchannels}
After the PDMS has been degassed and is ready for use, it is time to cut the microchannels from the PDMS mold. Follow these steps for precision and care in this process:

\begin{enumerate}
  \item Carefully peel the PDMS off the silicon wafer, ensuring that you do not damage the delicate microchannels. Use a gentle and steady hand to avoid any tears or distortions.

  \item Place the PDMS on a clean and flat surface. Ensure that the surface is free from any dust or debris that might affect the cutting process.

  \item With a very sharp blade, exacto knife, or scalpel, begin cutting the PDMS chips along the desired microchannel patterns. Take your time, keeping safety in mind and make precise cuts to ensure accuracy in the final microfluidic chip.

\end{enumerate}

By following these steps, you can successfully cut the PDMS microchannels from the mold, getting one step closer to the final microfluidic chip.

%\section{Creating inlets and outlets holes on the PDMS microchannels}
%\begin{figure}[ht!]
%\centering
%\includegraphics[angle=0,width=3.2in]{hole_punching.png}
%\caption{Hole punching in the microchip using a biopsy punch.}
%\label{holepunch}
%\end{figure}
%Place the PDMS on a clean and hard surface (preferably a cutting mat).
%Use the Biopsy puncher with 0.75mm diameter to punch four holes in the microchannels
%reservoirs.

\section{Creating Inlets and Outlets Holes on the PDMS Microchannels}
To complete the microfluidic chip, we need to create inlets and outlets for fluid flow. Follow these steps carefully:

\begin{figure}[ht!]
\centering
\includegraphics[angle=0,width=3.2in]{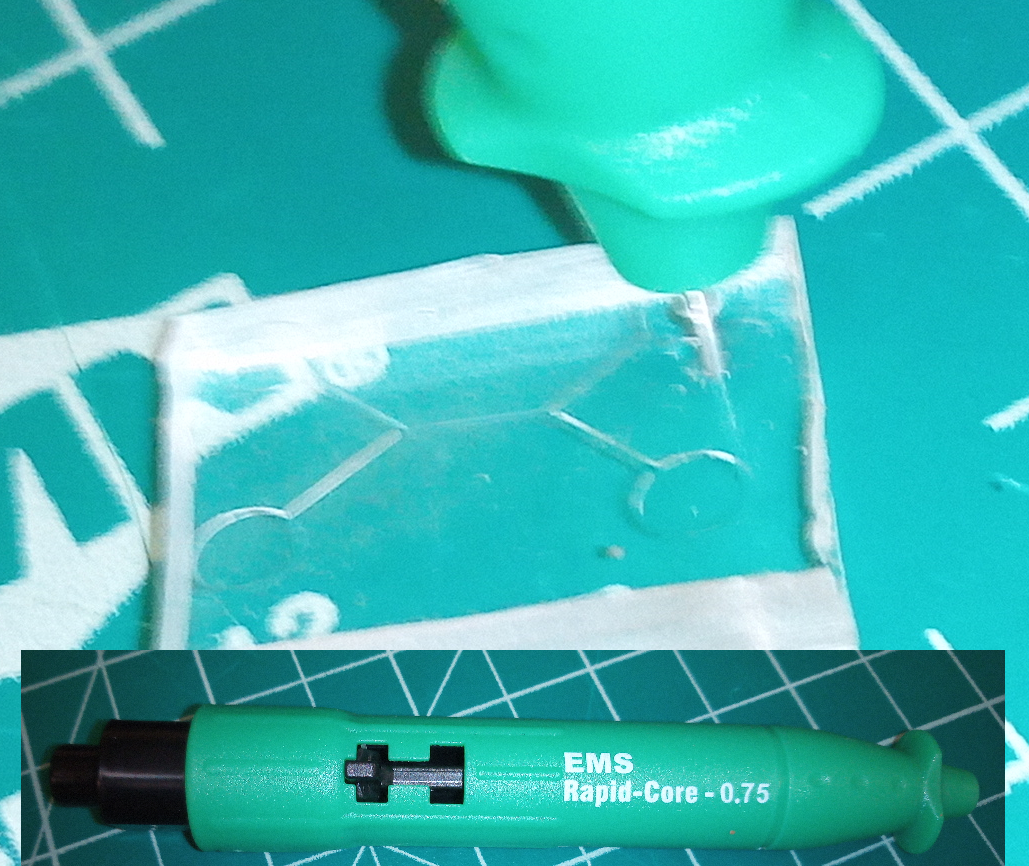}
\caption{Hole punching in the microchip using a biopsy punch.}
\label{holepunch}
\end{figure}

\begin{enumerate}
  \item Place the PDMS on a clean and hard surface, preferably a cutting mat, to ensure stability during the hole-punching process.

  \item Use the Biopsy puncher with desired diameter (we used a 0.75mm diameter puncher) to punch four holes in the microchannels' reservoirs. These holes will serve as the inlets and outlets for fluid introduction and extraction.

  \item Position the puncher carefully to align the holes precisely at the desired locations on the PDMS microchannels.

\end{enumerate}

By following these steps, you will successfully create the inlets and outlets holes on the PDMS microchannels, allowing for controlled fluid flow and efficient experimentation.

%\section{Bonding the PDMS microchannels with glass}

\section{Bonding the PDMS Microchannels with Glass}
To create a sealed microfluidic chip, we will bond the PDMS microchannels with a glass slide.

\begin{figure}[ht!]
\centering
\includegraphics[angle=0,width=5.0in]{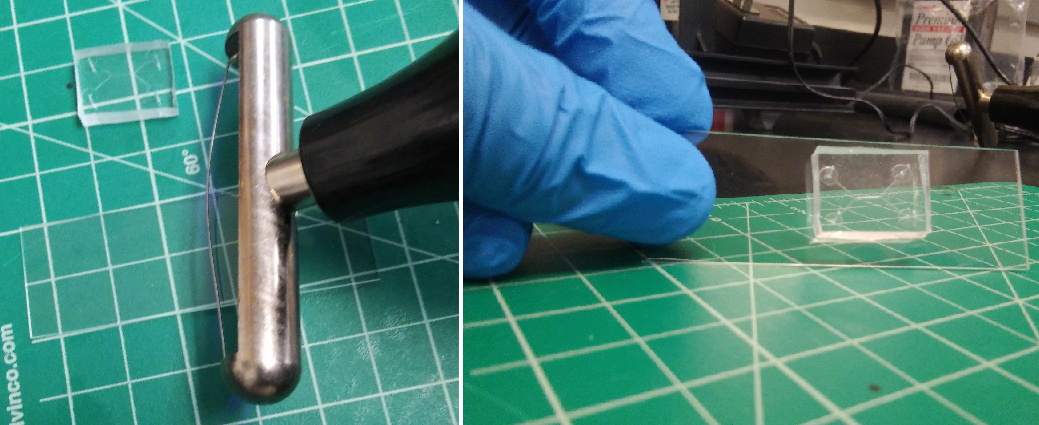}
\caption{Bonding the PDMS channel with glass.}
\label{Bonding}
\end{figure}

To irreversible bond PDMS channels with glass follow these steps carefully:

\begin{enumerate}
  \item Rinse the face of the PDMS with the microchannels using acetone, followed by isopropyl alcohol (IPA), and finally rinse with deionized (DI) water.

  \item Dry the PDMS thoroughly using clean compressed air to ensure there is no moisture left on the surface.

  \item Place the PDMS on a clean surface, ready for bonding.

  \item Take a clean glass slide and place it near the PDMS, ensuring both surfaces are free from any contaminants.

  \item Assemble and use the corona treater as defined in the instruction manual.

  \item Treat the face of the PDMS with the open microchannels using the corona treater for 30-60 seconds. This treatment enhances bonding properties.

  \item Treat one side of the glass slide with the corona treater for 30-60 seconds. We have also treated it with our lab generated microwave plasma \cite{barnes2021plasma} to get high quality bonding.

  \item Immediately bring together the treated surfaces of the PDMS and the glass slide, ensuring proper alignment.

  \item Apply gentle pressure on the PDMS and glass slide to bond them irreversibly. The corona treatment enhances adhesion between the materials.

  \item Put the bonded PDMS and glass slide in the incubator at 40$^{\circ}$C for 30 minutes. This step further strengthens the bond between the PDMS and glass slide.

\end{enumerate}

Following these steps, your microfluidic chip with bonded PDMS microchannels and glass slide is now ready for use.

Alternatively, if you have a plasma cleaner available in the lab, you can use it instead of the corona treater. Simply treat the glass and PDMS surfaces with plasma and bond them together following the same procedure.

%Rinse the face of the PDMS with the microchannels with acetone, IPA and DI water.
%Dry the PDMS with clean compressed air.
%Place the PDMS on a clean surface.
%Take a clean glass slide and place it near the PDMS.
%Take out the corona treater, assemble and use it as defined in the instruction’s manual.
%Treat with the corona treater for 30-60 seconds the PDMS face with the open microchannels.
%Treat one side of the glass slide with the corona treater for 30-60 seconds.
%Immediately bind both treated surfaces of the PDMS and glass slide.
%Apply some pressure gently on the PDMS and glass slide to bond them irreversibly.
%Put the bonded PDMS and glass slide in the incubator at 40$^{\deg}$C for 30min (this will strengthen
%the bond). The microchannel chip is now ready.
%Alternatively, if you have a plasma cleaner in the lab, use the plasma cleaner instead of the corona treater to treat the glass and PDMS surfaces with plasma and bond them together.
\section{Fitting the microchannel chip with the microfluidic system
L-shape connectors}
\begin{figure}[ht!]
\centering
\includegraphics[angle=0,width=5.0in]{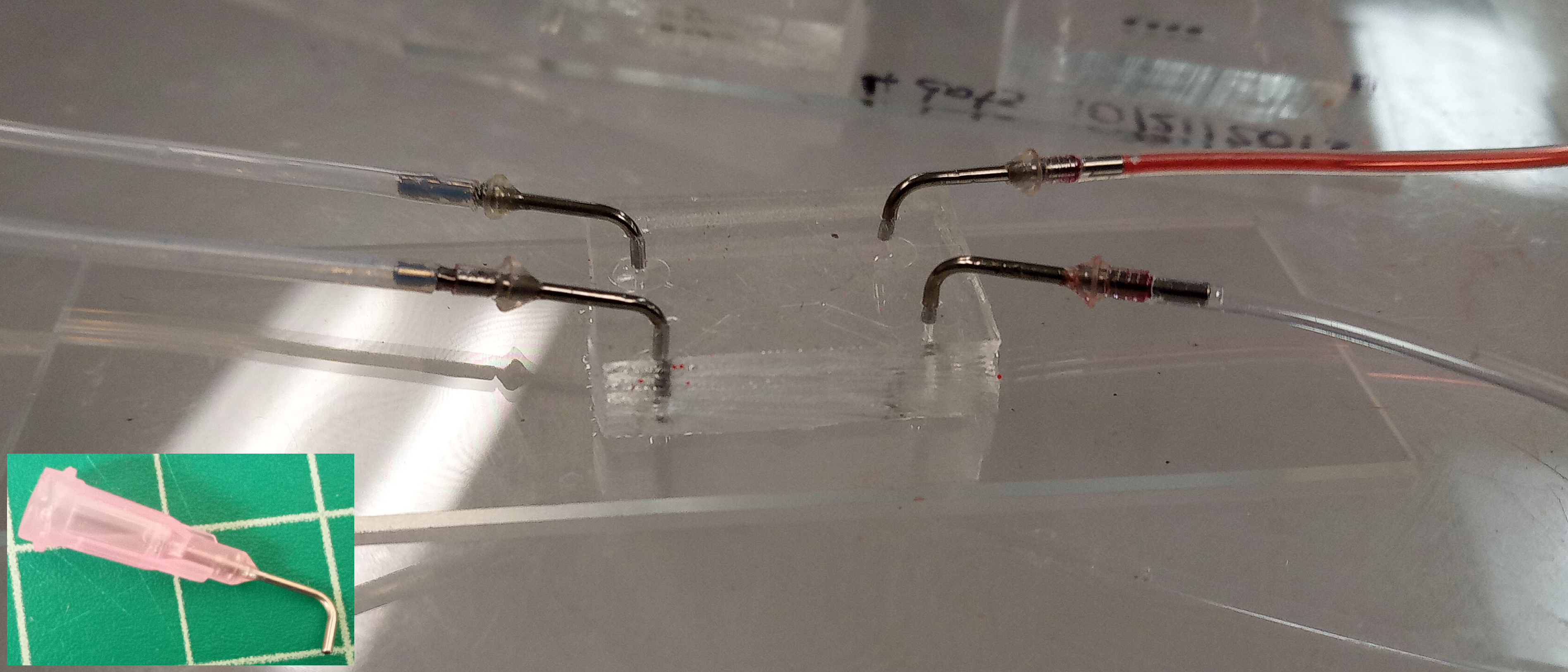}
\caption{Final fitting of the connectors and the tubes.}
\label{fitting}
\end{figure}
%Take the L-shaped syringe needle and clip out the plastic part. The diameter of the needle should be slightly more than the diameter of the biopsy puncher. In our case we had used 20 gauge needle (diameter 0.81 mm) which is slightle bigger than the PDMS holes (diameter 0.75 mm) to make the tubes fit the holes tightly to prevent any leaking.
%Connect one end of the L-shape connector to the tube connected to the syringe.
%Insert the other end of the L-shape connector inside the holes in the PDMS.
%Check that the L-shape connector is well-fitted inside the hole to avoid leaks.

To prepare the setup for fluid injection and extraction in the microfluidic chip, follow these steps:

\begin{enumerate}
  \item Take the L-shaped syringe needle and carefully remove the plastic part, leaving only the metal needle. The diameter of the needle should be slightly larger than the diameter of the biopsy puncher. In our case, we used a 20 gauge needle with an outer diameter of 0.81 mm, which is slightly larger than the PDMS holes (diameter 0.75 mm). This ensures a tight fit of the tubes in the holes, preventing any potential leaking.

  \item Connect one end of the L-shaped connector to the tube that will be connected to the syringe, ensuring a secure connection.

  \item Insert the other end of the L-shaped connector into the holes in the PDMS. Ensure that the L-shape connector is well-fitted inside the hole to avoid any leaks during fluid flow.

  \item Double-check the connections and make sure everything is securely in place before proceeding with any fluid injection or extraction.

\end{enumerate}

By following these steps, you can set up the fluid delivery system for your microfluidic chip, ensuring precise and controlled fluid manipulation.

\section{Experimental setup to test microfluidic mixing}
\begin{figure}[ht!]
\centering
\includegraphics[angle=0,width=5.0in]{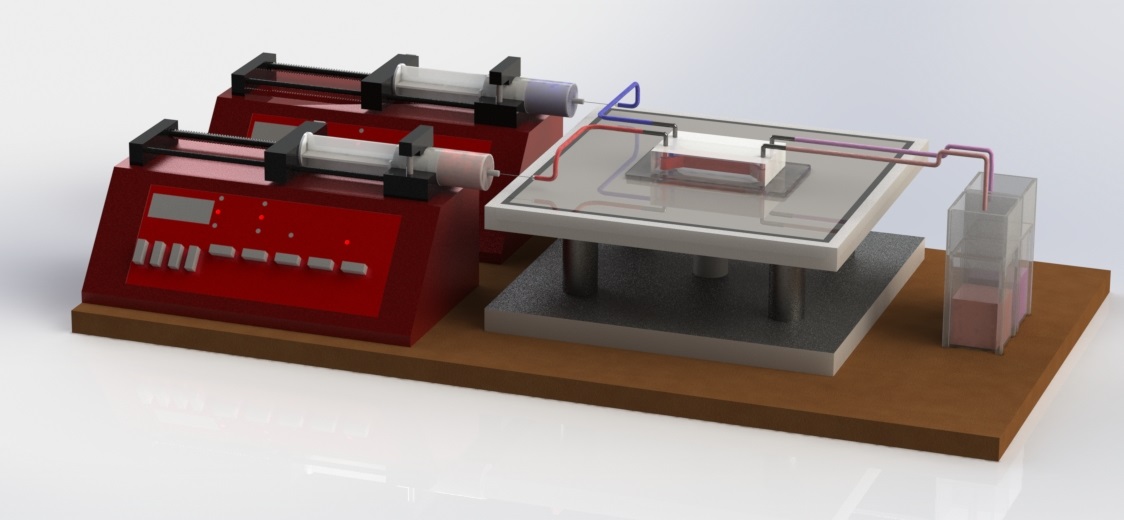}
\caption{Schematic of the microfluidic mixing experiment. One of the syringe pumps pushes a colored fluid and the other brings plain water to the microfluidic chip. Fluid is collected at the outlets  after going through the mixer and analyzed by a UV-VIS spectrometers.}
\label{schematic1pump}
\end{figure}

Fig.\ref{schematic1pump} shows the schematic of our fluid mixing experiment using the microfluidic chip created using the process described above in our undergraduate lab. Two miscible fluids were pumped into the microfluidic chip through the inlets by two syringe pumps and collected in the cuvettes placed at the outlet. The fluids collected at the outlet were analyzed by UV-VIS absorption spectroscopy to determine the mixing index and the overall effectiveness of the hydrophobic spots in the mixing of component fluids. In our experiment we used deionized (DI) distilled water (Barnstead NANOpureDiamond water purification system, specific resistivity 18.2 M$\Omega$-cm) in one of the syringe pumps, and DI water mixed with Allura Red dye (D=3.23$\times10^{-10}$m$^2$/s) in the other. Deltails of the experiment and the results can be found here \cite{ouro2022boundary}.

\section{Cleanup}
After the process, it is important to clean up the equipment used. Use acetone to wash off the Photoresist splashes on the spin coater. Use IPA to clean the tabletops used. Use the lab cleaning procedure to clean the glassware used and the tweezers.

\section{Acknowledgement}
KD would like to acknowledge Kavli Institute of Theoretical Physics (KITP) for their support through the KITP Fellowship program. This support facilitated KD's visit and stay at KITP, enabling the opportunity for potential collaborations. This work was supported in part by the National Science Foundation under Grant No. NSF PHY-1748958, the NSF HBCU-UP Award \# 1719425, and the Department of Education (MSEIP Award \# P120A70068) with MSEIP CCEM Supplemental grant.

\section{References}
\bibliography{BibtexdatabaseKSD}
\end{document}